\documentclass{emulateapj}\usepackage{apjfonts}

\newcommand\etal{{et~al.}} 
\newcommand\mM{\ifmmode(m{-}M)\else$(m{-}M)$\fi}

\newcommand\hst{{\it HST}}
\newcommand\zacs{\ifmmode z_{850}\else$z_{850}$\fi}
\newcommand\iacs{\ifmmode i_{775}\else$i_{775}$\fi}
\newcommand\gacs{\ifmmode g_{475}\else$g_{475}$\fi}
\newcommand\racs{\ifmmode r_{625}\else$r_{625}$\fi}
\newcommand\vacs{\ifmmode V_{606}\else$V_{606}$\fi}
\newcommand\feh{\ensuremath{[\hbox{Fe/H}]}}
\newcommand\feho{\ensuremath{\langle[\hbox{Fe/H}]\rangle_0}}
\newcommand\fehmax{\ensuremath{\langle[\hbox{Fe/H}]\rangle_{\rm max}}}
\newcommand\gz{{\ifmmode{(g{-}z)}\else$(g{-}z)$\fi}}
\newcommand\gzacs{{\ifmmode{g_{475}{-}z_{850}}\else$g_{475}{-}z_{850}$\fi}}
\newcommand\riacs{{\ifmmode{r_{625}{-}i_{775}}\else$r_{625}{-}i_{775}$\fi}}
\newcommand\rzacs{{\ifmmode{r_{625}{-}z_{850}}\else$r_{625}{-}z_{850}$\fi}}
\newcommand\izacs{{\ifmmode{i_{775}{-}z_{850}}\else$i_{775}{-}z_{850}$\fi}}
\newcommand\vzacs{{\ifmmode{V_{606}{-}z_{850}}\else$V_{606}{-}z_{850}$\fi}}
\newcommand\vi{{\ifmmode{(V{-}I)}\else$(V{-}I)$\fi}}
\newcommand\br{{\ifmmode{(B{-}R)}\else$(B{-}R)$\fi}}
\newcommand\gi{{\ifmmode{(g{-}i)}\else$(g{-}i)$\fi}}
\newcommand\bi{{\ifmmode{(B{-}I)}\else$(B{-}I)$\fi}}
\newcommand\zdf{$Z$DF}
\newcommand\zdfs{$Z$DFs}

\newcommand\siglf{\ensuremath{\sigma_{\rm LF}}}
\newcommand\sigfeh{\ensuremath{\sigma_{[{\rm Fe/H}]}}}

\newcommand\mbari{\ifmmode\overline{m}_I\else$\overline{m}_I$\fi}
\newcommand\mbarz{\ifmmode\overline{m}_z\else$\overline{m}_z$\fi}
\newcommand\mbar{\ifmmode\overline{m}\else$\overline{m}$\fi}

\newcommand\Mbar{\ifmmode\overline{M}\else$\overline{M}$\fi}
\newcommand\lbar{\ifmmode\overline{L}\else$\overline{L}$\fi}
\newcommand\Mbarz{\ifmmode\overline{M_z}\else$\overline{M}_z$\fi}
\newcommand\lta{\lesssim}
\newcommand\gta{\gtrsim}
\newcommand\cote{C{\^ o}t{\' e}}
\newcommand\jordan{Jord{\'a}n}

\shortauthors{{Blakeslee et al.}}
\shorttitle{Mass-Color-Metallicity Relations}
\slugcomment{To Appear in ApJ, Vol.\,710, February 2010}

\begin{document}

\title{The Mass-Metallicity Relation of Globular Clusters in the Context of Nonlinear Color-Metallicty Relations}

\author{John P. Blakeslee\altaffilmark{1},
Michele Cantiello\altaffilmark{2},
and
Eric W. Peng\altaffilmark{3,4}
}

\altaffiltext{1}{Herzberg Institute of Astrophysics, National Research Council of Canada, Victoria, BC V9E\,2E7, Canada; John.Blakeslee@nrc.ca}
\altaffiltext{2}{INAF-Osservatorio Astronomico di Teramo, via M. Maggini, I-64100, Teramo, Italy}
\altaffiltext{3}{Department of Astronomy, Peking University, Beijing 100871, China}
\altaffiltext{4}{Kavli Institute for Astronomy and Astrophysics, Beijing 100871, China}

\begin{abstract}
Two recent empirical developments in the study of extragalactic globular cluster
(GC) populations are the color--magnitude relation of the blue GCs (the ``blue
tilt'') and the nonlinearity of the dependence of optical GC colors on metallicity.  The
color--magnitude relation, interpreted as a mass--metallicity relation, is
thought to be a consequence of self-enrichment.  Nonlinear color--metallicity
relations have been shown to produce bimodal color distributions from
unimodal metallicity distributions.  We simulate GC populations including both a
mass--metallicity scaling relation and nonlinear color--metallicity relations
motivated by theory and observations.  Depending on the assumed range of metallicities 
and the width of the GC lumin\-osity function (GCLF), we find that the
simulated populations can have bimodal color distributions with a ``blue~tilt''
similar to observations, even though the metallicity distribution appears unimodal.
The models that produce these features have the relatively high mean GC
metallicities and nearly equal blue and red peaks characteristic of giant elliptical
galaxies.  The blue tilt is less apparent in the models with metallicities typical
of dwarf ellipticals; the narrower GCLF in these galaxies has an even bigger effect
in reducing the significance of their color-magnitude slopes.
We critically examine the evidence for nonlinearity versus bimodal metallicities as
explanations for the characteristic double-peaked color histograms of
giant ellipticals and conclude that the question remains open.
%
We discuss the prospects for further 
theoretical and observational progress in constraining the models presented here and
for uncovering the true metallicity distributions of extragalactic
globular cluster systems.
\end{abstract}


\section{Introduction}
\label{sec:intro}

Massive early-type galaxies have globular cluster (GC) systems that follow bimodal
color distributions (e.g., Gebhardt \& Kissler-Patig 1999; Larsen \etal\ 2001;
Harris \etal\ 2006; Peng \etal\ 2006, hereafter P06).  Before this was a commonly
recognized feature of such systems, Ashman \& Zepf (1992) discussed the
possibility of GC color bimodality based on the simple idea that ellipticals are
the remnants of dissipationally merged pairs of spirals (e.g., Toomre 1977).  In
this picture, a merger-formed population of red, relatively high-metallicity GCs
would then appear distinct from the blue, metal-poor GCs of the progenitor
spirals.  

The actual formation histories of ellipticals appear to have been more complex
(see for example the review by Renzini 2006).  In particular, most of the stars in large
ellipticals today are believed to have formed early, in a series of rapid
starbursts at redshifts $z>2$, whereas the hierarchical assembly of these galaxies
continued to $z<1$ through ``dry'' (dissipationless) mergers (e.g., De Lucia
\etal\ 2006; Menci \etal\ 2008; Kodama \etal\ 2007; Thomas \etal\ 2005).  Such
hierarchical formation models do not naturally produce bimodal GC metallicity
distributions in most elliptical galaxies (Beasley \etal\ 2002; Kravtsov \& Gnedin
2005).  Instead, the stochastic nature of the formation would be expected to
result in approximately Gaussian distributions.  Thus, while some galaxies with
quiet histories like the Milky Way's may have bimodal GC metallicity distributions
if they formed their stars in just a few discrete bursts, it is difficult to see
how bimodality would be nearly universal in this context.

An alternative model was put forth by \cote\ \etal\ (1998) in which ellipticals
\textit{form} with unimodal, relatively rich GC metallicity distributions and
later acquire the blue GC component through dissipationless mergers with
early-type dwarfs.  This explanation appears more consistent with hierarchical
formation models.  However, the universality of the bimodality, even at
intermediate luminosities, and the asymmetric tail of red GCs in dwarf
ellipticals, would again be surprising in this scenario.

Models of GC color distributions have typically assumed linear conversions
between metallicity and optical colors.  The observed color bimodality 
thereby became equated with bimodality in metallicity.  In contrast, P06
showed that the empirical transformation from metallicity to the
broad-baseline \gz\ color is distinctly nonlinear.  An earlier indication of
such nonlinearity for the $(C{-}T_1)$ color index (Harris \& Harris 2002; Cohen
\etal\ 2003) has also been confirmed (Lee \etal\ 2008). 
Further, Yoon et al.\ (2006)
pointed out that nonlinear color-metallicity relations predicted by their simple
stellar population (SSP) models naturally produce bimodal color histograms from
unimodal metallicity distributions.  Such nonlinearities are commonly predicted
by realistic SSP models that try to reproduce the structure in stellar color-magnitude
diagrams as a function of metallicity (see Cantiello \& Blakeslee
2007, hereafter CB07).  This provides an alternative, simple explanation for the
near-universality of bimodal GC color distributions, although the idea
remains controversial as we discuss in Sec.~\ref{sec:discussion} below.

It is not surprising that as observations and models improve, deviations from the
simplest linear assumptions become important.  A related example is the
dependence of galaxy surface brightness fluctuation magnitudes on integrated color:
linear fits sufficed at the level of accuracy of ground-based data, but the latest
\hst\ studies (Mei \etal\ 2007; Blakeslee \etal\ 2009) clearly show a wavy relation
reminiscent of that between metallicity and color. 
Galaxy scaling relations are another example where nonlinearity has become
evident amid the continuity of structural properties in new samples of galaxies
spanning many orders of magnitude (\cote\ \etal\ 2008).
Even the Cepheid period-luminosity relation has been found to exhibit significant
nonlinearity (Ngeow \etal\ 2009).  Although the errors in derived quantities
(e.g., distance modulus or metallicity) resulting from such nonlinearities are
often small for individual objects, the implications for the distributions of
these quantities derived for large populations can be profound.

Another development has been the discovery that GCs in the blue peak of the color
distribution tend to be redder at brighter magnitudes (Harris \etal\ 2006; Mieske
\etal\ 2006; Strader \etal\ 2006).  Dubbed ``the blue tilt,'' the effect is
subtle, but further evidence confirms it is real (Harris 2009; Peng \etal\ 2009).
The likely explanation is that the most massive GCs were able to self-enrich to a
small extent, thus producing a mass-metallicity relation (e.g., Dopita \& Smith
1986; Morgan \& Lake 1989; Brown \etal\ 1991; Recchi \& Danziger 2005; Strader \&
Smith 2008; Bailin \& Harris 2009).  The tendency is much less obvious for the
red-peak GCs, despite the otherwise lockstep behavior of the two peaks, which is
strong evidence that they formed under similar conditions (P06; Larsen \etal\
2001).  However, the red peak is broader because of the nonlinear
color-metallicity relation, and it usually does not reach as high in optical
luminosity, at least partly because of the decrease in luminosity-to-mass ratio
with increasing metallicity.  A weak ``red tilt'' is expected theoretically
(Bailin \& Harris 2009), and there is evidence for this in some cases (Mieske
\etal\ 2006; Lee \etal\ 2008).

The purpose of this work is to bring together these two recent developments in the
study of extragalactic GC systems: we explore the nature of the blue tilt
in the context of unimodal metallicity distributions with nonlinear
dependence of color on metallicity.  To do this, we assume a mass-metallicity
relation consistent with the findings of recent studies.  The following section
describes our approach in more detail, and Sec.~\ref{sec:results} presents the
consequences for the observed blue tilt under various assumptions about the range
of GC metallicities and luminosities.
The implications of our results and the question of bimodality are
considered in Sec.~\ref{sec:discussion}.  The final section summarizes our conclusions.

\section{Method}
\label{sec:method}

Our goal is to investigate the presence and significance of the ``blue tilt''
--- the color-magnitude relation of the blue peak GCs --- in the case of 
unimodal metallicity distributions following an empirically motivated
mass-metallicity relation (MMR) and the nonlinear color-metallicity relations
both observed and predicted by detailed SSP modeling.  We also examine the 
``tilt'' of the red peak GCs in this scenario.  For the sake of
generality in this initial investigation, we try to keep our assumptions as
simple as possible while still being realistic enough to yield useful
insights.

\begin{figure}\epsscale{1}
\plotone{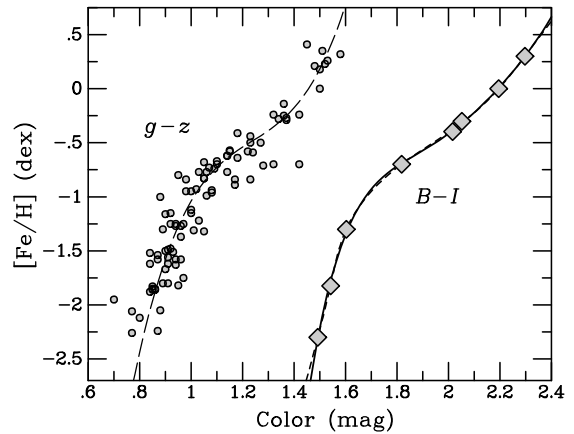}
\caption{Empirical \gz\ and model \bi\ color-metallicity relations.  The circles
  show data for Milky Way and Virgo cluster GCs (Peng \etal\ 2006), and the
  long-dashed curve is a fourth-order polynomial fit to the data.  The
  diamonds show predictions for \bi\ from the 13 Gyr Teramo SPoT models used in
  this study. The solid curve is a fifth-order polynomial fit, while the
  short-dashed curve is a cubic spline interpolation/extrapolation of the models.
  As commonly used for these indices, the \bi\ colors
  are on the Vega-based Johnson-Cousins system while \gz\ is given in AB magnitudes.
  The conversion from the AB to Vega system for \gz\ is $+$0.64~mag;
  thus, these color indices are quite similar, but the data have a steeper slope
  than the models at the red end.
\label{fig:nonlinear}}
\vspace{1cm}
\end{figure}

\subsection{Color--Metallicity Relations}
\label{ssec:color-met}

P06 presented an empirical \feh-\gz\ relation based on their photometry and
spectroscopic metallicities for a combination of 95 Milky Way, M49 (NGC\,4472), 
and M87 GCs
(see their Fig.~11).  The Milky Way GC data are part of a larger project
to produce a homogeneous database of integrated colors for the entire 
Galactic GC system in the SDSS photometric bandpasses (West \etal\ 2006;
M.~West \etal, in preparation). 
The inclusion of the Virgo galaxy GC metallicities from Cohen \etal\ (1998, 2003) was necessary to
characterize the relation at high metallicity.   This adds uncertainty
because the conversion from the Lick indices to the Zinn \& West (1984)
metallicity scale is not well calibrated above solar metallicity and involves
some extrapolation (see Fig.~14 of Cohen \etal\ 1998).  However, the
extrapolation is strictly linear and extends $\lta0.5$~dex at the metal-rich
end, so it could not introduce a nonlinearity in the \feh-\gz\ relation.  
Intrinsic trends in age
and/or alpha-enhancement with metallicity could potentially introduce biases in
the metallicity scale (see the discussion in P06), but
such trends would not change our results if they involve a simple scaling
factor.  Since we wish to avoid ad~hoc assumptions about model ages and
abundance ratios as far as possible, we prefer
to adopt empirical calibrations in our analysis, and
this remains the best empirically-based color-metallicity relation available.

For simplicity, P06 used a broken linear fit to the curved \feh-\gz\ relation,
excluding the highest metallicity GCs which would have required additional curvature.
To better characterize the full data set presented in P06, we fit a quartic
polynomial using robust orthogonal regression (GaussFit;
Jefferys \etal\ 1988) and find the following relation
\begin{eqnarray}
\hbox{[Fe/H]} \;=\; -33.74 \,+\, 79.81\gz \,-\, 66.69\gz^2 \nonumber \\
\,+\, 20.66\gz^3  \,-\, 1.08\gz^4 \,,
\label{eq:empirical}
\end{eqnarray}
which we plot in Figure~\ref{fig:nonlinear} along with the data points from P06.
Note that the empirical \gz\ colors are calibrated to the AB system, and 
the conversion from AB to Vega-based magnitudes for \gz\ is $+$0.64~mag.

Other color indices used in recent studies of GC color
distributions and MMRs include \bi\ (e.g., Harris \etal\ 2006; Harris 2009),
\vi\ (Peng \etal\ 2009), \br\ (Spitler \etal\ 2006; Cantiello \etal\ 2007),
$(g-i)$ (Wehner \etal\ 2008), and
$(C-T_1)$ (Forte \etal\ 2007; Lee \etal\ 2008).  
The empirical metallicity dependencies of these indices have not been
characterized in the same detail as \gz; usually the conversions to metallicity
are based on simple linear approximations.  However, CB07 present 
color-metallicity predictions in the $UBVRIJHK$ bandpasses from the Teramo 
``SPoT'' models\footnote{http://www.oao-teramo.inaf.it/SPoT} (Raimondo \etal\ 2005)
and make comparisons to other SSP models in the context of
color-metallicity relations.  
Figure~\ref{fig:nonlinear} also shows the color-metallicity predictions from the
13~Gyr SPoT models for \bi, the color index which first revealed the presence of
the MMR in the GCs of giant ellipticals (Harris \etal\ 2006), along with simple
polynomial and cubic spline fits to the models.  (The age of 13~Gyr is the same as
for the models used by Yoon \etal\ [2006] and CB07.)  The plotted \gz\ and \bi\
relations provide empirical and theoretical examples of nonlinear
color-metallicity relations for use in this study.  See Yoon \etal\ (2006) for a
theoretical \gz-metallicity relation and CB07 for predicted relations in other
bandpasses.

\begin{figure}\epsscale{0.9}
\plotone{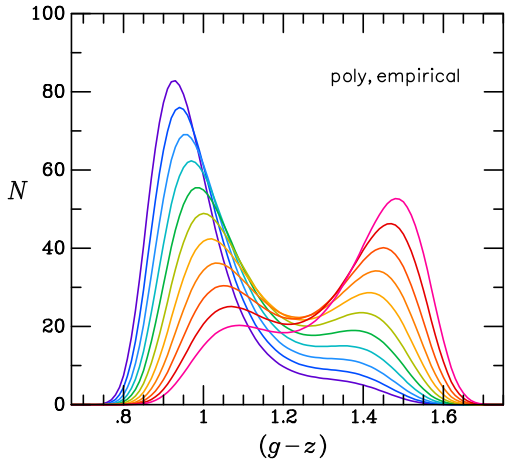}
\caption{Calculated \gz\ color distributions based on the empirical polynomial fit
  in Fig.~\ref{fig:nonlinear} for a series of unimodal Gaussian \feh\
  distributions having a common dispersion of 0.5~dex and means
  ranging from $-1.2$ to $-0.2$~dex in steps of 0.1~dex.  
  The normalization of the curves is arbitrary; the color ranges
  from violet/blue at low metallicity to red/magenta at high metallicity.
Although the mean color increases with metallicity and the peaks shift 
redward, the position of the trough remains fairly constant near the point where
the derivative of the metallicity with respect to color has a minimum.
\label{fig:gzhistos}}
\vspace{0.1cm}
\end{figure}

\begin{figure}\epsscale{1.17}
\hspace{-0.3cm}
\plottwo{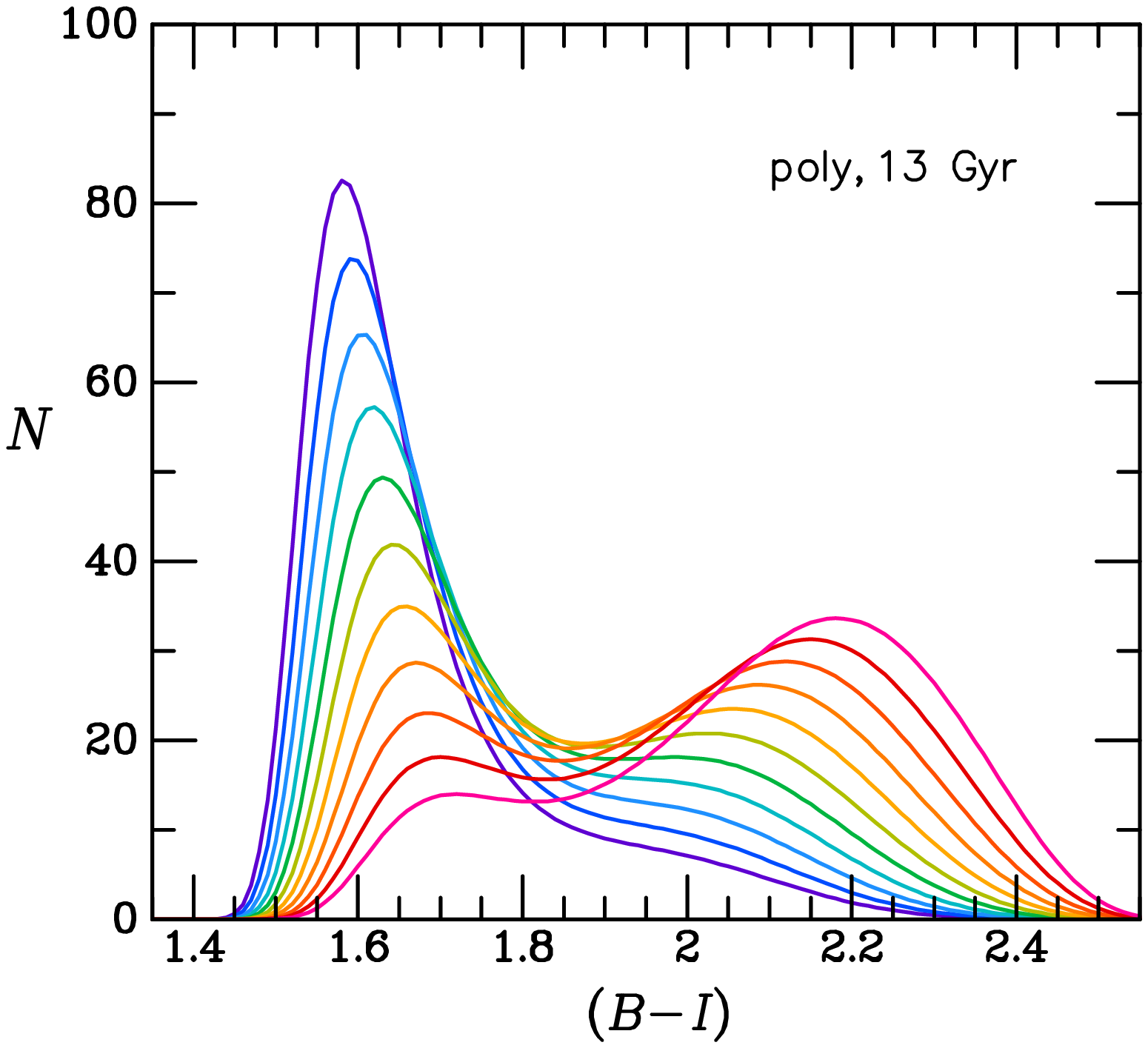}{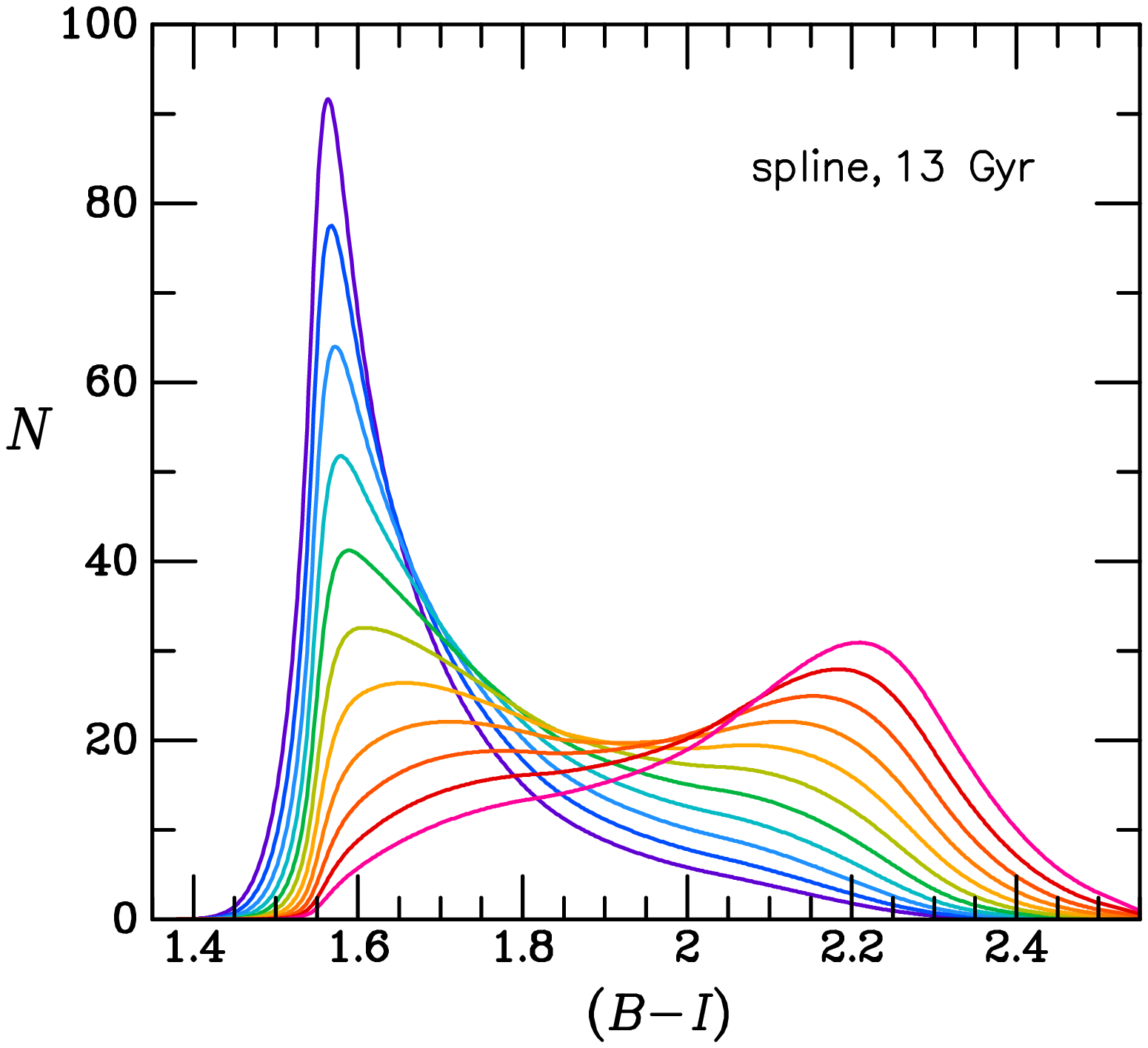}
\caption{\bi\ color distributions given by the polynomial (left) and
  spline (right) fits to the model points in Fig.~\ref{fig:nonlinear}
  for a series of unimodal Gaussian \feh\
  distributions having a common dispersion of 0.5~dex and means
  ranging from $-1.2$ to $-0.2$~dex in steps of 0.1~dex.  
  The normalization of the curves is arbitrary; the color ranges
  from violet/blue at low metallicity to red/magenta at high metallicity.
  Although the polynomial and spline fits to the model points in
  Fig.~\ref{fig:nonlinear} are very similar, the derived color distributions
  depend on their derivatives and thus show marked differences.  
  For real populations, observational errors would largely obscure such
  differences. 
\label{fig:bihistos}}
\vspace{0.16cm}
\end{figure}

\subsection{Mass--Metallicity Scaling}
\label{ssec:massmet}

There is evidence that the MMR is already present near the peak of the GC
luminosity function (GCLF), although it is more prominent a couple magnitudes
brighter than this (e.g., Harris \etal\ 2006; Mieske \etal\ 2006; Peng \etal\
2009).  Using linear color-metallicity relations for blue GCs, these studies
derive scaling relations between GC luminosity $L$ and metallicity $Z$ consistent
with $Z\sim L^{0.5}$ for massive elliptical galaxies (e.g., Cockcroft \etal\
2009; Strader \& Smith 2008), or a slope of $-$0.2~dex~mag$^{-1}$ for \feh\ (we
confine the discussion here to solar-scaled models).  This coefficient is not
observationally well constrained, as the galaxy-to-galaxy scatter is large and the
applied color-metallicity conversions are approximate.

We implement this MMR scaling by assuming that the
$Z$ distribution function (\zdf) is a Gaussian in the logarithm with $\sigfeh = 0.5$
dex (e.g., Yoon et al.\ 2006; CB07) at each magnitude, but with the mean \feh\
increasing by 0.2 dex per magnitude brighter than the GCLF peak, up to some 
limit (see Sec.~\ref{ssec:params}). 
Figure~\ref{fig:gzhistos} shows the predicted \gz\ color
distributions for Gaussian \zdfs\ with $\sigfeh = 0.5$ dex and means varying
from $-$1.2 to $-$0.2~dex.  Consistent with the observations (e.g., P06),
even at low mean metallicities there is a substantial ``red tail'' to the
color distribution; moreover, the distribution always appears
distinctly bimodal at the highest metallicities appropriate to GC systems, i.e.,
the blue peak is always present.  

Figure~\ref{fig:bihistos} presents the corresponding plots for \bi\ based on the
polynomial and spline fits to the SPoT model predictions.  We show both sets of
curves to illustrate the fact that since the color distributions depend on the
derivative of the nonlinear relation between metallicity and color, the 
interpolation method can affect the details of the derived distributions.  In
practice, observational errors would obscure most of these details.  For the
analysis below, we concentrate on the results derived using the polynomial
interpolation to be more consistent with the approach used for the empirical \gz\
relation (for which spline interpolation is not a reasonable option).  However, we
have carried out the analysis using both methods, and the results are 
qualitatively the same.  In this case, the uncertainties and limitations of the
model predictions 
are greater than the differences due to the interpolation scheme.

The GCs in the simulations follow a standard Gaussian GCLF at any
given metallicity.
Since the MMR is most prominent in giant ellipticals, we use a GCLF width $\siglf
= 1.4$ mag, characteristic of the most luminous galaxies (e.g., Harris 1991;
Blakeslee \etal\ 1997; Harris \etal\ 2009), except in the discussion of the MMR
for dwarf ellipticals in Sec.~\ref{ssec:dwarfs} below.  We apply
the model mass-to-light values to obtain the correct relative luminosities at
different metallicities, so that the overall population follows a single
Gaussian distribution in mass (e.g., Ashman \etal\ 1995).

\subsection{Monte Carlo Simulations}
\label{ssec:params}

The main adjustable parameters in this analysis are the mean \feh\ at the peak
of the GCLF, which we label \feho, and the maximum mean metallicity that we allow
for the brightest GCs, which we label \fehmax.  In other words, we 
impose a fixed mean of \fehmax\ for the \zdf\ of the GCs above some luminosity
limit, rather than continuing the scaling relation indeterminately.
The mean \feh\ of the GC system could be derived from \feho,
\fehmax, and the GCLF.  Empirically, the mean GC metallicity increases with galaxy
luminosity (\cote\ \etal\ 1998; P06).  We do not explicitly include galaxy
properties in these simulations, but we consider a large enough range in \feho\
and \fehmax\ to make useful comparisons to galaxies with a range of luminosities.

Given the above set of assumptions, we generate random realizations of GC
systems and examine their color distributions and color-magnitude relations.
We simulate realistic observational scatter by adding random magnitude errors
that scale exponentially as $\sigma_m\sim10^{-0.4\,m}$, and we set the 
normalization such that the color error at the GCLF peak is 0.1~mag.  We also
include an additional 0.01~mag scatter at all magnitudes, which could
represent errors in aperture corrections or intrinsic scatter in the GC
colors at fixed metallicity.  
We experimented with larger values of these errors, but we note that
the blue tilt is only detectable with the best available photometric data and
analyses, and we do not wish to obscure the effect we are trying
to characterize.

\subsection{Color-Magnitude Fitting Procedure}
\label{ssec:kmm}

To characterize the blue (and red) tilts in our simulated samples, we basically
follow the procedure described in Mieske \etal\ (2006), which is similar to that
used in other studies (e.g., Peng \etal\ 2009).  Specifically, we bin the GCs by
magnitude, using a minimum bin size of 0.2~mag but allowing the size in magnitudes
to increase in order to contain at least 200 GCs per bin.  We then use the
heteroscedastic mode of the KMM mixture modeling algorithm (Ashman \etal\ 1994) to
determine the peaks of the color distributions for the different bins.  We 
perform linear fits to determine how the peak positions vary as a function of
magnitude. The fits are done using all bins that are more than 1$\,\siglf$
brighter than GCLF turnover magnitude; for giant ellipticals this is close to the 
``million solar mass'' threshold (Harris 2009) where the tilt becomes noticeable in
observational data. The following section presents the results for the color-magnitude
slopes for the blue and red peaks in these simulations.

\begin{figure}\epsscale{1.11}
\vspace{0.1cm}
\plotone{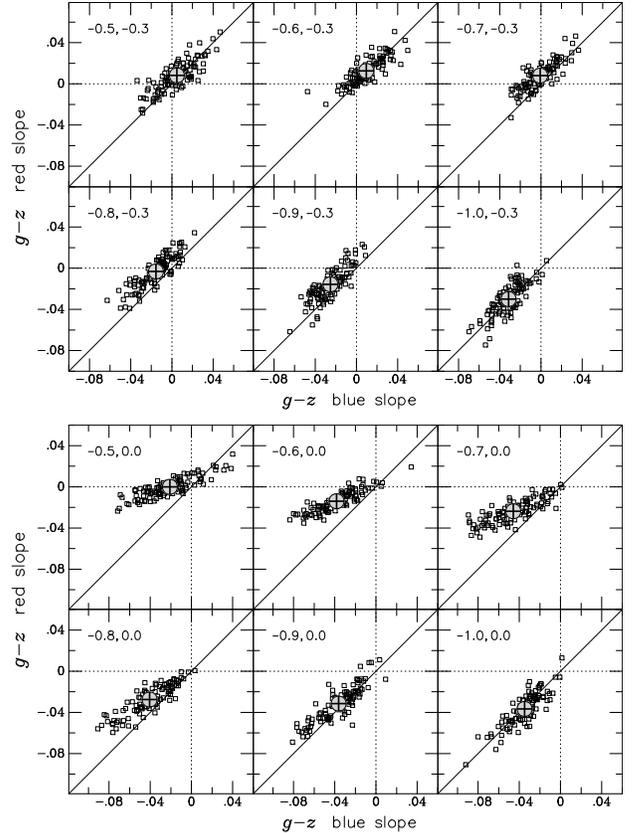}
\caption{Blue versus red tilts in simulated GC \gz\ color distributions.
Each point represents the slopes of the blue and red peak colors determined by KMM
fits as a function of magnitude within a single realization of a population of
10,000~GCs.  See text for details on KMM fitting.
Each panel shows the slopes for 101 different realizations (squares), all with
the same model parameters.  The large encircled crosses mark the average values.
From panel to panel, we vary two parameters:
\feho, which is the mean metallicity for GCs at the mean luminosity of the GCLF, 
and \fehmax, which is mean metallicity of the most luminous GCs 
in the population.  The values of these parameters in dex are displayed 
in the upper left of each panel; for reasons given in the text,
we show results for only two values of \fehmax.
At any given luminosity, the GCs follow a unimodal metallicity
distribution of width  $\sigfeh = 0.5$ dex.  
Here we use $\siglf = 1.4$ mag for the GCLF width, appropriate for massive
early-type galaxies.  We note that several of these panels show a strong
preference for ``blue tilts,'' i.e., negative blue-peak slopes that are
significantly larger in absolute value than the red-peak slopes.
\label{fig:gzslopes_giant}}
\vspace{0.15cm}
\end{figure}

\begin{figure}\epsscale{1.1}
\plotone{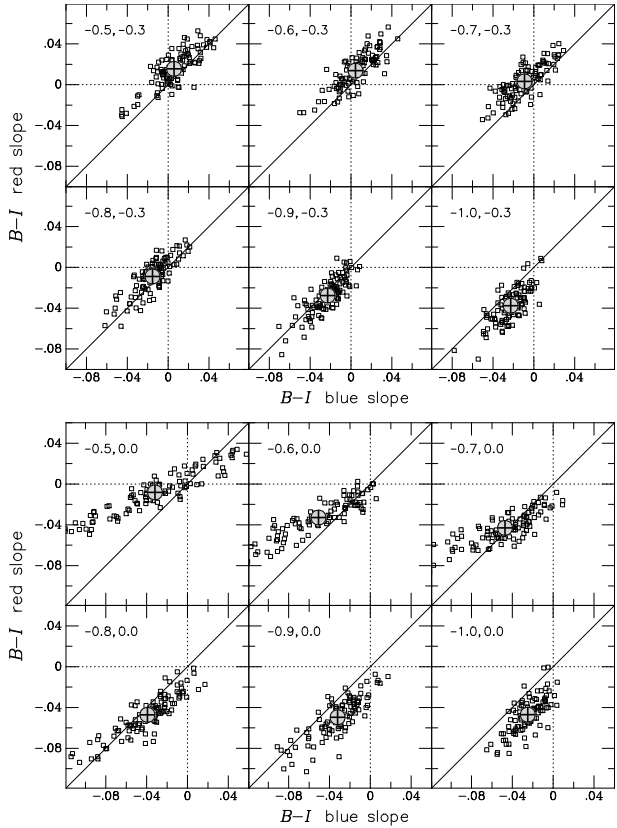}
\caption{Same as Fig.~\ref{fig:gzslopes_giant}, but here shown for 
the slopes with mag of the fitted KMM blue and red peaks in 
the simulated GC \bi\ color distributions.
\label{fig:BIslopes_giant}}
\end{figure}

\section{Results}
\label{sec:results}

\subsection{Blue vs Red Tilts for Baseline Models}
\label{ssec:baseline}

Figure~\ref{fig:gzslopes_giant} plots the slope of the blue peak in the \gz\ color
distribution determined by KMM as a function of magnitude versus the corresponding slope
for the red peak.  Each panel shows results for 101 random realizations of a GC
populations with 10,000 members brighter than the GCLF peak.  The twelve different
panels are for different combinations of \feho\ and \fehmax, defined in
Sec.~\ref{ssec:params}; their values are shown in each panel.  In particular, the
mean metallicity at the GCLF peak \feho\ is varied 
from $-1.0$~dex (one-tenth solar) 
to $-0.5$~dex (one-third solar),
but only two values, $-0.3$ (half solar) and 0.0 (solar), are shown for the
maximum mean metallicity parameter \fehmax.  Lower values of \fehmax\ decrease the
mean metallicity range of the MMR to factors of only $\sim\,$2 and/or restrict the
simulated populations to lower mean metallicities and bluer mean colors than those
found in the giant ellipticals where the ``blue tilt'' is commonly observed.
Setting \fehmax\ to values above solar metallicity (or removing it altogether)
gives results similar to the case with $\fehmax=0.0$~dex, already a high
metallicity for GCs.  Thus these two values of \fehmax\ prove adequate for our
purposes.

The points in Figure~\ref{fig:gzslopes_giant} 
with negative slope values are the ones that show
a reddening of the peak colors with increasing luminosity; thus, these are the
populations that would empirically be described as having a positive MMR.
The third row of panels in Figure~\ref{fig:gzslopes_giant} is particularly
interesting, as the blue-peak slopes are
generally larger in absolute value than the red-peak slopes.
%
For example, in the three panels of Figure~\ref{fig:gzslopes_giant} with $\fehmax=0.0$
and $\feho = -0.5$, $-0.6$, $-0.7$ dex, the mean \gz\ blue-peak slopes are
$\langle{d\gz/dz}\rangle_{B} = -0.023$, $-0.039$, $-0.046$, respectively, and the
red-peak slopes are $\langle{d\gz/dz}\rangle_{R} = -0.001$, $-0.014$, $-0.024$.  The
scatter in these slope values is typically $\sim0.02$, so the uncertainties on the
mean slopes are about $\pm0.002$.
If a linear transformation were used to convert the slopes from color to metallicity,
the conclusion would be that the blue GCs have a stronger MMR, but this is not the
case here, where there is a single MMR and no bimodality in metallicity.
For comparison, analysis of a combination ACS Virgo and Fornax Cluster Survey
(\cote\ \etal\ 2004; \jordan\ \etal\ 2007a) galaxies gives mean \gz\ blue and red
peak slopes of $-0.035\pm0.008$ and $-0.016\pm0.010$, respectively, with
substantial scatter among different subsamples (Mieske \etal\ 2010).
Thus, the simulation with $\feho,\fehmax = -0.6,0.0$ dex provides a
very good match to the observed \gz\ blue and red tilts.

Even for the top two rows of panels in Figure~\ref{fig:gzslopes_giant}, there is a
general preference for blue tilts over red ones.  More specifically, where both
blue and red slopes are negative, there is a tendency for the blue slope to be
more negative, that is, to have a more pronounced color-magnitude relation.  For
example, the simulation with $\feho,\fehmax = -0.9,-0.3$ (second row, middle
panel) has mean \gz\ blue and red slopes of $-0.025\pm0.002$ and $-0.016\pm0.002$,
respectively.  We choose this as a lower metallicity comparison because it has the
same factor-of-four range in the MMR limits and also favors the blue tilt, though
to a lesser degree.

\begin{figure}\epsscale{1.17}
\vspace{0.1cm}\hspace{-0.25cm}
\plotone{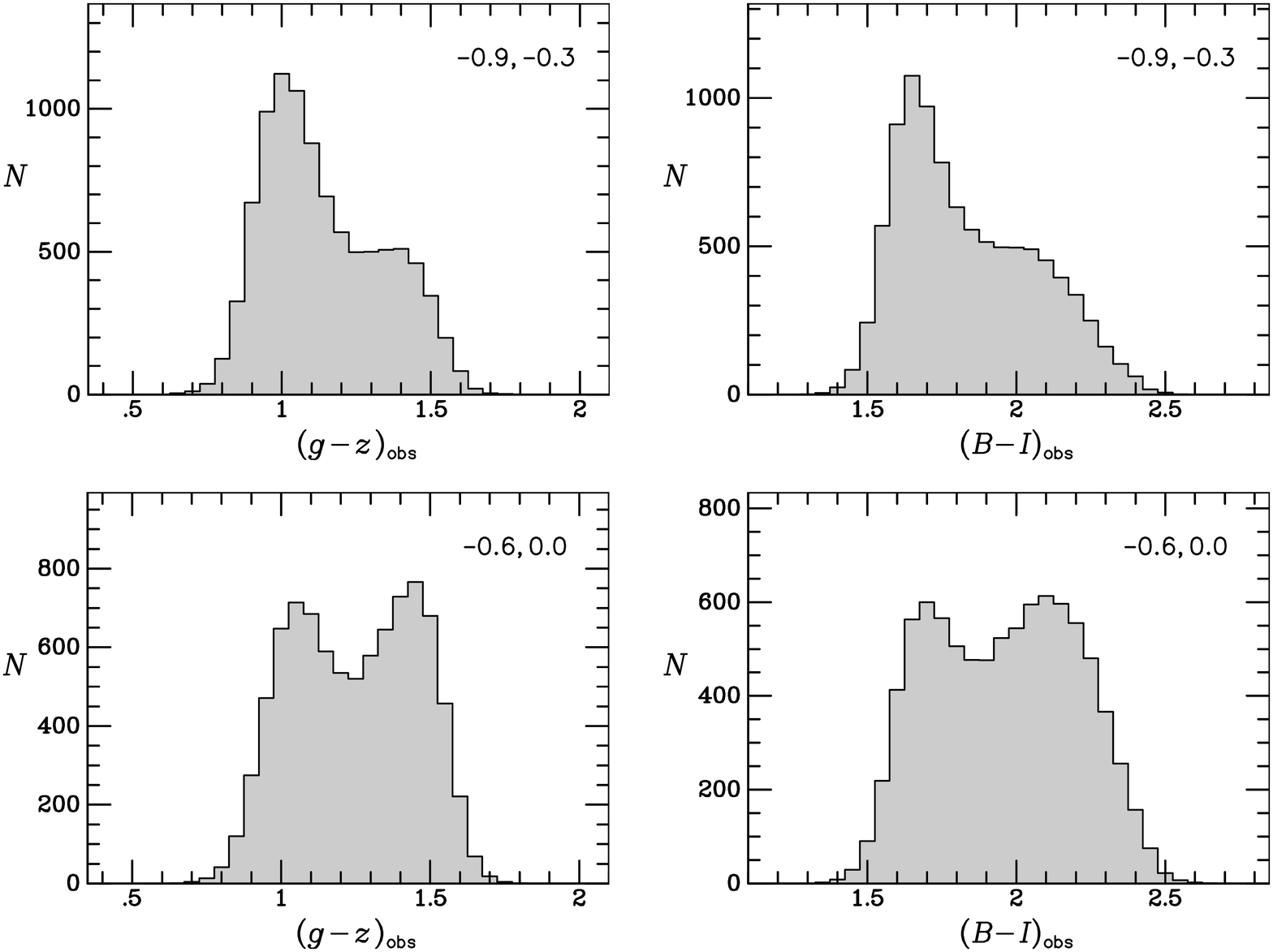}
\caption{Example histograms of \gz\ and \bi\ colors for GCs in two
  simulated populations with values of \feho\ and \fehmax\ given in
  the upper right of each panel.  
The top panels are typical of the GC color distributions for intermediate
luminosity galaxies, while the lower panels resemble the distributions for the
most massive early-type galaxies (e.g., Peng \etal\ 2006).  Both of
  these simulations show a tendency towards blue tilts for the \gz\
  colors; only the higher metallicity model tends towards blue tilts
  for the \bi\ colors, which are derived from a model, rather than
  empirical, color-metallicity relation.
\label{fig:color_histos}}
\end{figure}

\begin{figure}\epsscale{1.0}
\plotone{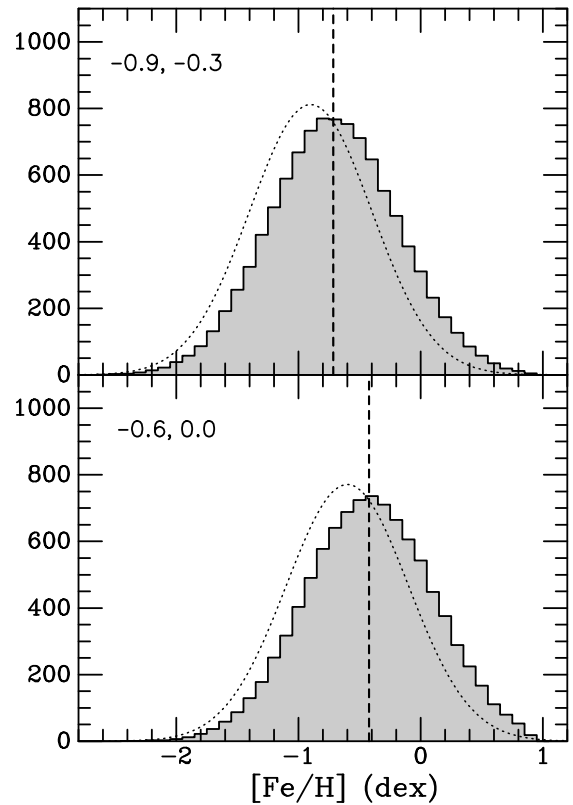}
\caption{Metallicity distributions (grey histograms)
 for the simulations whose color histograms
  are displayed in Figure~\ref{fig:color_histos}.  We plot metallicities only for
  model GCs brighter than the GCLF peak, since there is no mass-metallicity
  scaling at fainter luminosities.  Including the fainter half of the
  GCLF would shift the peak value lower by $\sim\,$0.08~dex.
  For comparison, the dotted curves show Gaussian metallicity
 distributions with mean = \feho, $\sigfeh=0.5$, and normalized to have the same total areas
 as the histograms.
  The dashed lines mark the mean metallicities
  of the plotted histograms.
\label{fig:metal_histos}}
\end{figure}

Figure~\ref{fig:BIslopes_giant} presents analogous results from the same set of
simulations but for the slopes of the blue and red peaks of the \bi\ color
distributions with magnitude.  As described above, the transformation from
metallicity to \bi\ colors is based on the 13~Gyr Teramo SPoT models, whereas the
\gz\ colors are from the P06 empirical transformation.  Overall, the results are
fairly similar, but the tendency to favor blue tilts is somewhat weaker among
these simulations using the model-based transformations.  In this case, the model
with $\feho,\fehmax = -0.6,0.0$ (second row, middle panel) has mean blue and red
slopes $\langle{d\bi/dI}\rangle_{B} = -0.049\pm0.003$ and
$\langle{d\bi/dI}\rangle_{R} = -0.032\pm0.002$.  The blue slope is very similar to
the value of $-0.05$ found by Harris (2009), but that study found no significant
red slope, although Bailin \& Harris (2009) state that it should present at some
level.  For the model with $\feho,\fehmax = -0.9,-0.3$, the blue and red slopes
are $-0.021$ and $-0.026$, respectively, so the red tilt is greater in this case,
although neither is very large.  Despite this, we consider it an interesting
lower-metallicity comparison model on account of the empirically-based \gz\ slopes
discussed above.

Figure~\ref{fig:color_histos} presents example \gz\ and \bi\ histograms for the
two models we have been discussing.  These histograms exemplify the range of color
distributions typically found in intermediate- to high-luminosity elliptical
galaxies.  The distributions are clearly bimodal in all the panels of this figure.
In the top panels the blue peak dominates, giving an appearance very similar to
the GC color distributions found by P06 for Virgo galaxies with
$M_B\approx-19$~mag.  On the other hand, the roughly equal peaks in the histograms
of the lower panels of Figure~\ref{fig:color_histos} are quite similar to those
found for the most massive early-type galaxies (e.g., P06; Harris \etal\ 2006).
These are the galaxies that also show the blue tilt most strongly.  It is
interesting that galaxies with relatively more dominant blue peaks would have
weaker blue tilts; this would be puzzling for a trivial linear conversion between
color and metallicity.  However, as shown by the slope-slope comparisons in
Figures~\ref{fig:gzslopes_giant} and~\ref{fig:BIslopes_giant}, the more realistic
color-metallicity conversions used here reproduce this observational finding.

\begin{figure}\epsscale{1.16}
\vspace{0.12cm}\hspace{-0.24cm}
\plottwo{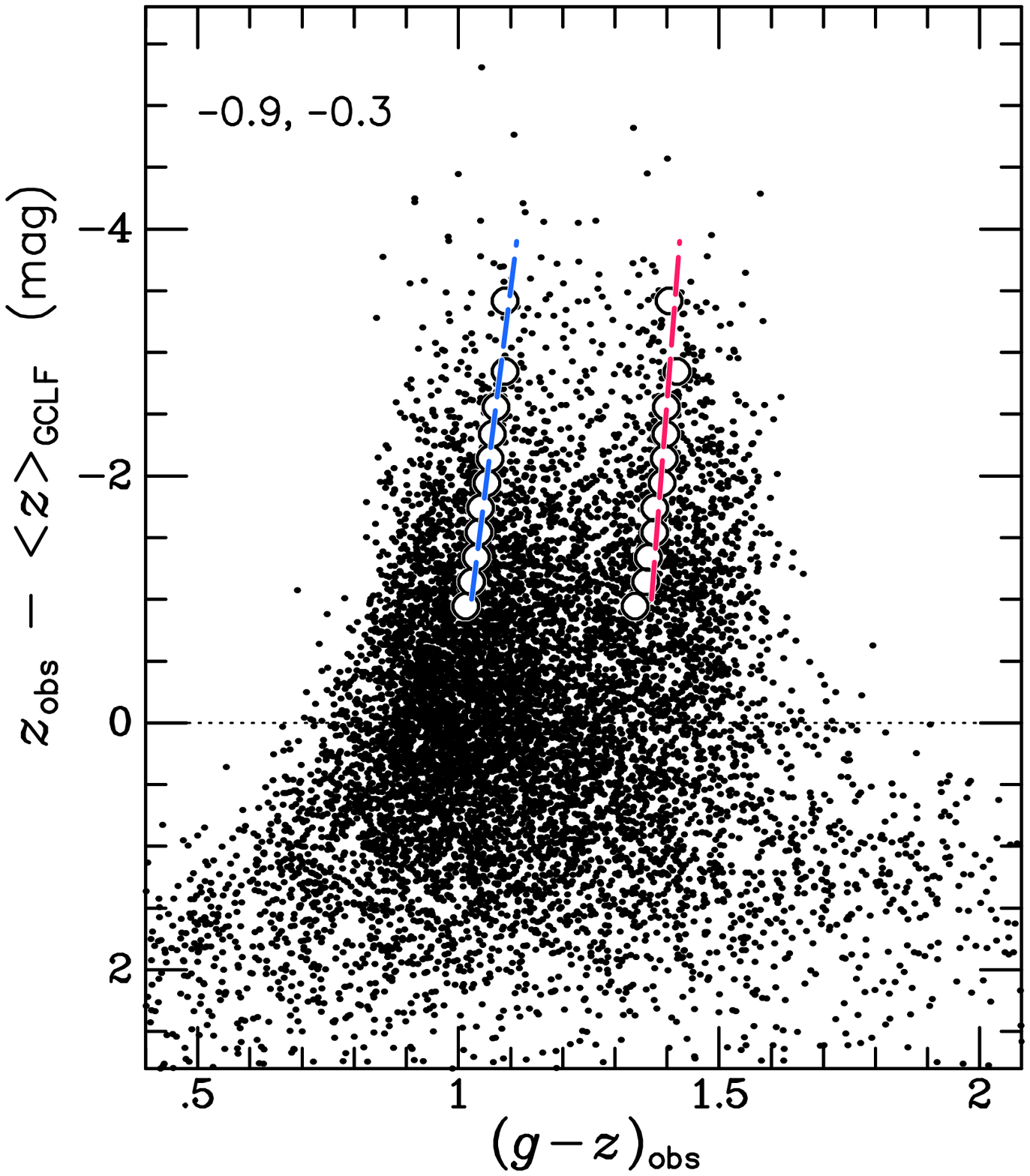}{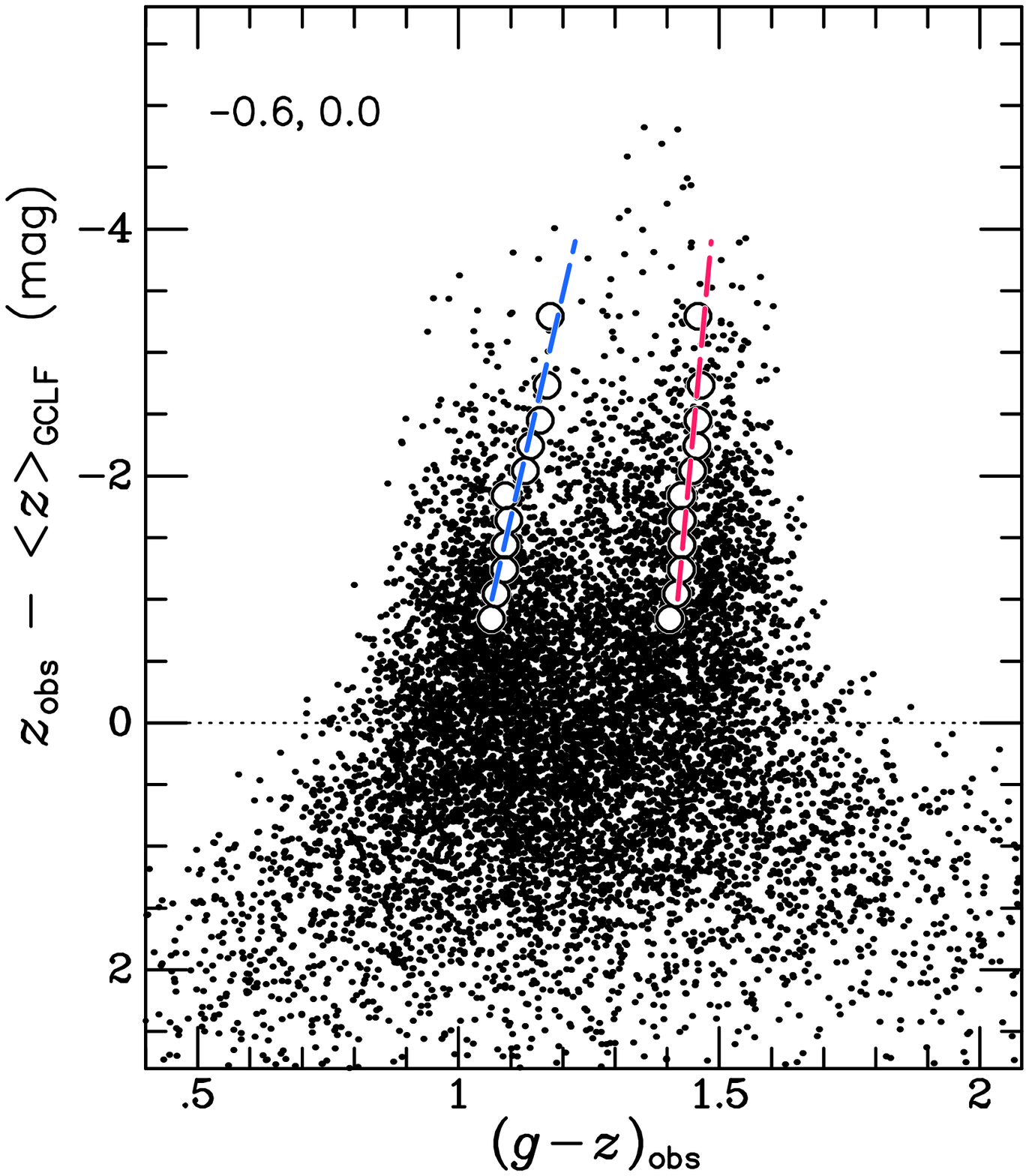}
\caption{%
Color-magnitude diagrams for two simulations discussed in the text.
The values of \feho\ and \fehmax\ are shown, and the magnitudes are given with
respect to the GCLF mean.  We plot points for 10,000 GCs in each
simulation.  The large circles show the results from the KMM fits, which are
based on 20~times the number of GCs as are plotted in these panels.  The dashed
blue and red lines are linear fits to the KMM peak results.
These diagrams are illustrative, but in real populations there may be
unmodeled correlations of mean GC metallicity and color with galaxy luminosity,
GCLF width, or other parameters. 
\label{fig:cmds}}
\vspace{0.25cm}
\end{figure}

In comparison to the color histograms, 
Figure~\ref{fig:metal_histos} shows that despite the MMR built into the models,
the underlying metallicity distributions still appear unimodal.  Thus, the
bifurcation in color results purely from the nonlinearity of the color-metallicity
conversion, not the varying of GC mean metallicity with luminosity.
Figure~\ref{fig:cmds} illustrates the $z$ versus \gz\ color-magnitude diagrams
for these baseline models.   For clarity, the plotted simulations contain 
10,000 GCs each.  The larger circles show results from KMM fits
as described in Sec.~\ref{ssec:kmm}.  In order to delineate the trends more
clearly, the KMM fits in Figure~\ref{fig:cmds} are based on 20~times the
number of GCs as are plotted in the panels. 
As with the color histograms, these color-magnitude diagrams bear a good
resemblance to those found observationally (e.g., P06, Mieske \etal\ 2006).

\begin{figure}\epsscale{1.1}
\vspace{0.1cm}
\plotone{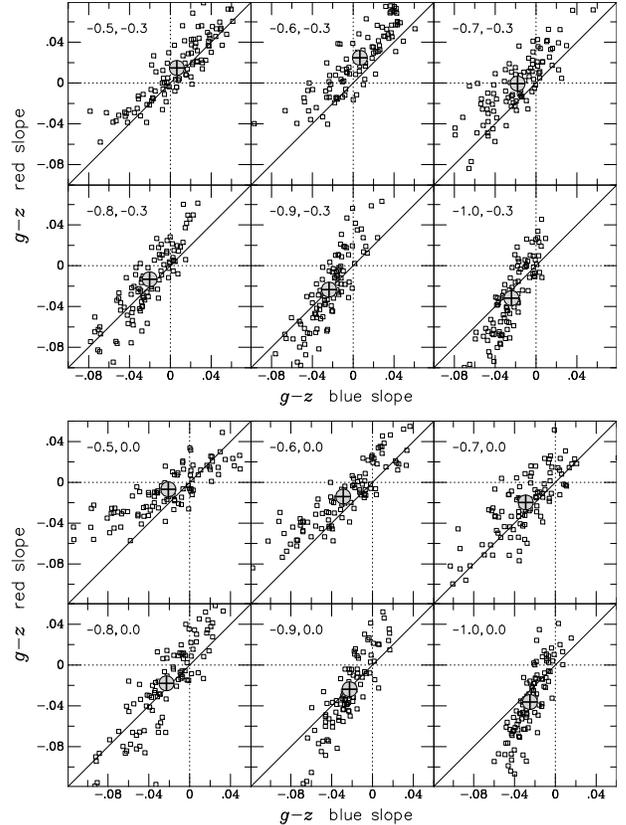}
\caption{%
Same as Fig.~\ref{fig:gzslopes_giant}, but here shown for sets of simulations
using $\siglf=0.9$~mag, typical of the GCLF widths of dwarf galaxies.  Although
some individual realizations may have large ``tilts'' (slopes), the scatter is
large and encompasses $(0,0)$ at the $\lta1\sigma$ level.  Thus, there is no
significant tendency for either red or blue tilts.
\label{fig:gzslopes_dwarf}}
\vspace{0.3cm}
\end{figure}


\vspace{0.1cm}
\subsection{Dwarf Galaxies: the Effect of the GCLF}
\label{ssec:dwarfs}

Observationally, the color-magnitude relation of blue-peak GCs is mainly a
property of giant ellipticals, and the trend is weak or absent in lower
luminosity galaxies (e.g. DeGraaff \etal\ 2007; Cantiello \etal\ 2007).
However, Mieske \etal\ (2006, 2010) have detected it in pooled samples of GCs from faint
early-type galaxies.  As we have shown, the simulations with lower
metallicity, which are more appropriate to low luminosity ellipticals, have less
of a tendency to favor the blue tilt than do the higher metallicity simulations.
However, the likely more important reason for the lack of strong blue tilts in dwarf
galaxies has to do with their GCLFs.  Not only do they have many fewer GCs, but their
luminosity functions are narrower.  If the GCLF is too narrow, the GCs will not 
reach high enough luminosities to evince a significant MMR.

\jordan\ \etal\ (2006) found that the width of the GCLF decreases linearly with
magnitude.  For early-type galaxies with $M_B\gta-18$, the width $\siglf\lta0.9$ mag; this includes the
majority of the ACS Virgo and Fornax Survey galaxies, which are actually dominated
by dwarf ellipticals.
Figure~\ref{fig:gzslopes_dwarf} shows results for the slopes with magnitude of the
peaks in the \gz\ color distributions analogous to
Figure~\ref{fig:gzslopes_giant}, but for populations with a narrower GCLF
$\siglf=0.9$ mag, appropriate to dwarf ellipticals.  The scatter in the slope
values is roughly a factor of two larger for the case with $\siglf=0.9$ as
compared to $\siglf=1.4$, and encompasses $(0,0)$ at the $\lta1\sigma$ level in all cases.
Measurements of the slopes in individual
realizations are generally not significant at more than the 1$\,\sigma$ level.
Thus, the lower metallicities and narrower GCLF can explain why the observed
GC color-magnitude trends are very weak in low luminosity galaxies.

\subsection{Age Issues}
\label{ssec:age}

\begin{figure}\epsscale{1.17}
\vspace{0.2cm}\hspace{-0.3cm}
\plottwo{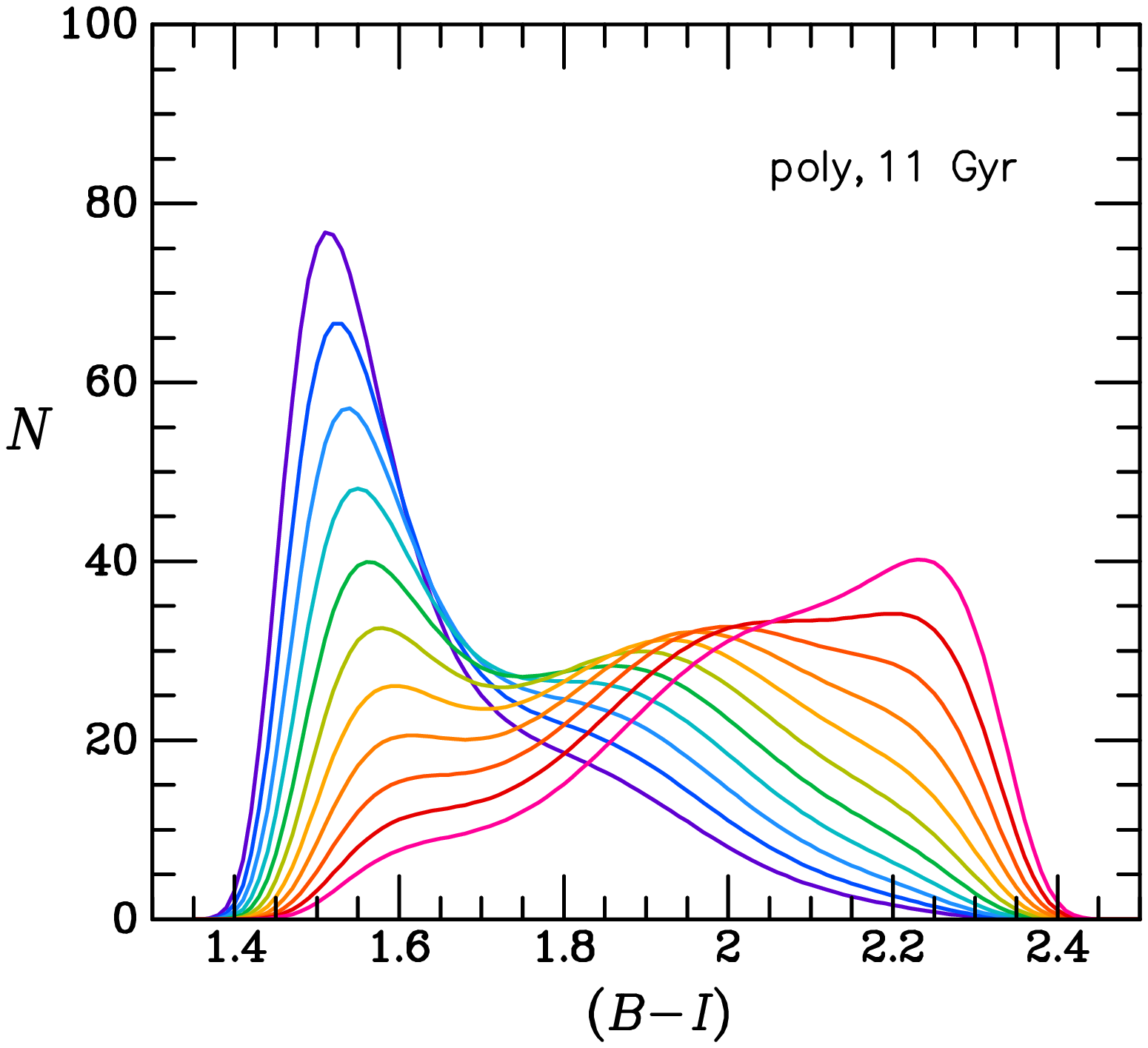}{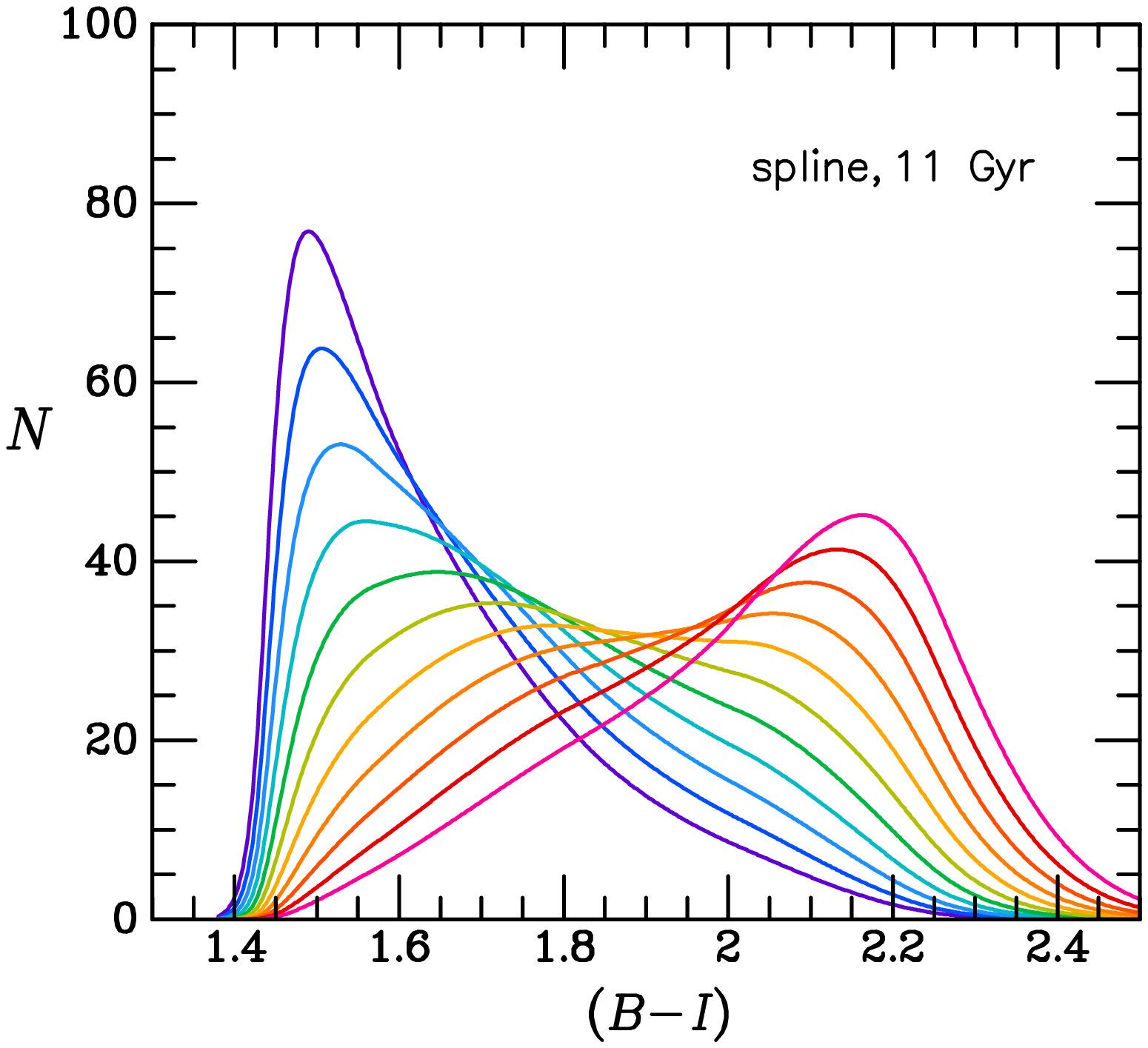}
\caption{Similar to Fig.~\ref{fig:bihistos},
  but here based on the color--metallicity predictions from the 
  11~Gyr SPoT models.  Compared to the 13~Gyr models, there are more significant
  differences in the color distributions derived here from the polynomial and
  spline fits.  A more complete set of model metallicities would help alleviate
  this interpolation uncertainty.
\label{fig:bi11gyrhistos}}
\vspace{0.4cm}
\end{figure}

So far the discussion has centered on GC color distributions from simulations
using the empirical \gz-metallicity relation and a theoretical \bi-metallicity
relation from the 13~Gyr SPoT models.  As shown previously in
Figures~\ref{fig:gzhistos} and~\ref{fig:bihistos}, both of these produce clearly
bimodal color distributions from single Gaussians in metallicity.  As another
example, Figure~\ref{fig:bi11gyrhistos} shows predicted \bi\ distributions
similar to those in Figure~\ref{fig:bihistos} but now based on the 11~Gyr SPoT
models.  The agreement between the distributions derived with the polynomial and
spline interpolations is worse here than for the 13~Gyr models, mainly because
the polynomial fit is poorer.
At the highest metallicities, the polynomial-derived curves in
Figure~\ref{fig:bi11gyrhistos} appear almost trimodal, whereas the spline-derived
curves appear fairly similar to those in Figure~\ref{fig:bihistos}, except
that the blue peak is a bit broader.

\begin{figure}\epsscale{1.12}
\plotone{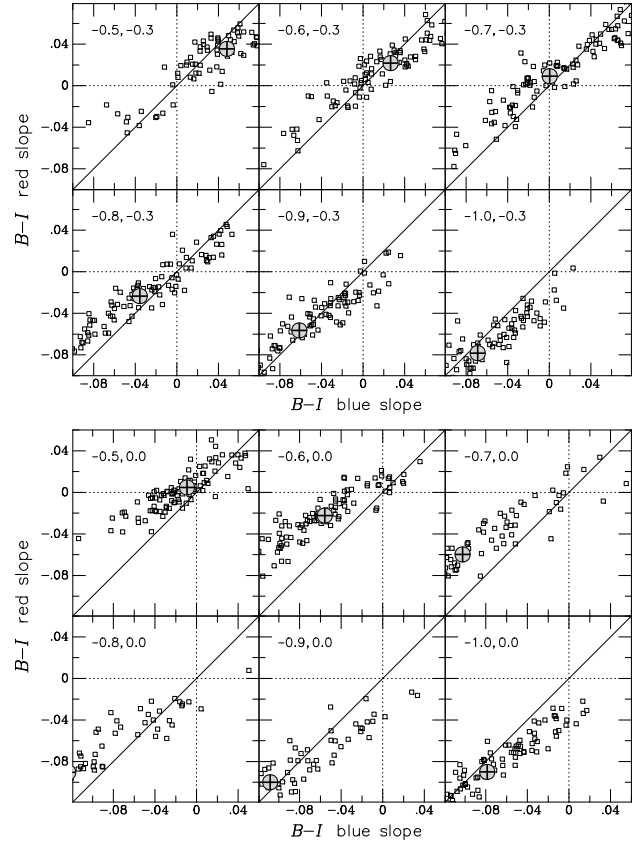}
\caption{%
Same as Fig.~\ref{fig:BIslopes_giant}, but here based on color-metallicity
relations from the 11~Gyr models (using the polynomial fits to the
color-metallicity relation, as in Fig.~\ref{fig:BIslopes_giant}).
\label{fig:BIslopes_11gyr}}
\vspace{0.35cm}
\end{figure}

To examine the effects of these more irregular color distributions, 
Figure~\ref{fig:BIslopes_11gyr} shows results for the slopes of the
peaks in the \bi\ color distributions analogous to
Figure~\ref{fig:BIslopes_giant}, but based on the polynomial fits to the
color-metallicity relations from the 11~Gyr SPoT models. 
The scatter is higher by at least $\sim\,$50\%, and in some cases several times larger,
because the distributions are less clearly bimodal,
making the KMM fits much less robust.  As in the case for the 13~Gyr models, there
is no clear preference for blue tilts in the
two rows of panels in Figure~\ref{fig:BIslopes_11gyr} with $\fehmax=-0.3$ dex.
However, there is a tendency towards blue tilts in the third row of panels, in the
sense that the slopes are negative and the blue slopes 
tend to be greater in absolute value than the red slopes.  In particular, for our
fiducial ``giant galaxy'' model with $\feho,\fehmax = -0.6,0.0$ dex,
the mean blue and red slopes are
$\langle{d\bi/dI}\rangle_{B} = -0.059\pm0.005$ and
$\langle{d\bi/dI}\rangle_{R} = -0.022\pm0.003$.
This indicates an even stronger blue tilt than for the 13~Gyr models, though again with
much larger scatter.  Unfortunately, the behavior of the color-metallicity
relation as a function of age is not well calibrated, and as we have seen, the
results in this case depend on the interpolation method.
Of course, the \gz\ results based on the empirical
color-metallicity transformation do not explicitly assume any age for the GCs, and
the simulated \gz\ distributions are strongly bimodal with a clear tendency for the blue tilt.

We further note that since the point of inflection of the nonlinear
color-metallicity relation shifts with age, it is also possible to introduce a
blue tilt by making the brightest GCs have systematically older ages (S.-J.~Yoon,
private communication).  However, there is no a~priori reason to assume a gradient
in age with luminosity, whereas the MMR that we have used in our simulations is
reasonable based on GC self-enrichment.  Related to this, it would be very useful
to have observations of GC color distributions at various lookback times to see
how the color bimodality changes with time.  If there is underlying bimodality in
metallicity, it should change very little.  On the other hand, if it is due to the
inflection in the color-metallicity relation, then the bimodality should decrease
even at modest lookback times of a few Gyr (e.g., compare
Figures~\ref{fig:bihistos} and~\ref{fig:bi11gyrhistos}).  
In practice, the observational requirements are forbidding, since it involves
measuring precise GC color distributions at $z\gta0.2$.

Kalirai \etal\ (2008) attempted to measure evolution in the GC colors
of a $z{=}0.089$ elliptical galaxy serendipitously
observed in extremely deep \hst\ imaging of a foreground Galactic globular
cluster.  These authors detected two peaks in the color distribution and
interpreted them as the usual local peaks, but shifted substantially
\textit{redder} at $z{=}0.089$ (even after extinction and $k$~corrections).
Large \textit{anti}-evolution in color at a lookback time of just over
1~Gyr would not be easy to under\-stand in the context of either
explanation for
color bimodality.  However, from Fig.~2 of Kalirai \etal, it appears that the
true blue and red peaks are probably where they should
be, but they are unresolved because of observational error at these faint magnitudes.  
The apparent red peak in the color
distribution is likely contamination from the coolest members of the white
dwarf sequence in the foreground cluster.  Thus, this study illustrates the
difficulty of such observations.  However, it would be worth trying this
experiment with the next generation of astronomical observatories, such as the
\textit{James Webb Space Telescope} (\textit{JWST}) and ground-based extremely
large telescopes (ELTs).  In the meantime, further refinements in the model
color-metallicity predictions for different ages and elemental abundance ratios
are greatly needed.

\vspace{1cm}
\section{Discussion: Bimodality or Nonlinearity?}
\label{sec:discussion}


The nonlinear dependence of GC optical colors on metallicity is found both empirically
and from detailed SSP modeling.  Because the observed and predicted relations naturally
produce bimodal color distributions, this nonlinearity is the simplest explanation for
the observed universality of bimodal color distributions.  The simplicity of this
explanation is the strongest argument in its favor.
In contrast, \textit{universal} bimodality in the underlying GC metallicity
distributions would be difficult to understand in the context of hierarchical formation
models.  However, this argument is purely a theoretical bias, and it is important to
critically examine the available data.  Of course, even if the nonlinearity is the
principal cause of the characteristic double-peaked GC color histograms, this would
not preclude a predominance of two distinct metallicity components in \textit{some}
galaxies.

\begin{figure}\epsscale{1.15}
\vspace{0.15cm}\hspace{-0.15cm}
\plottwo{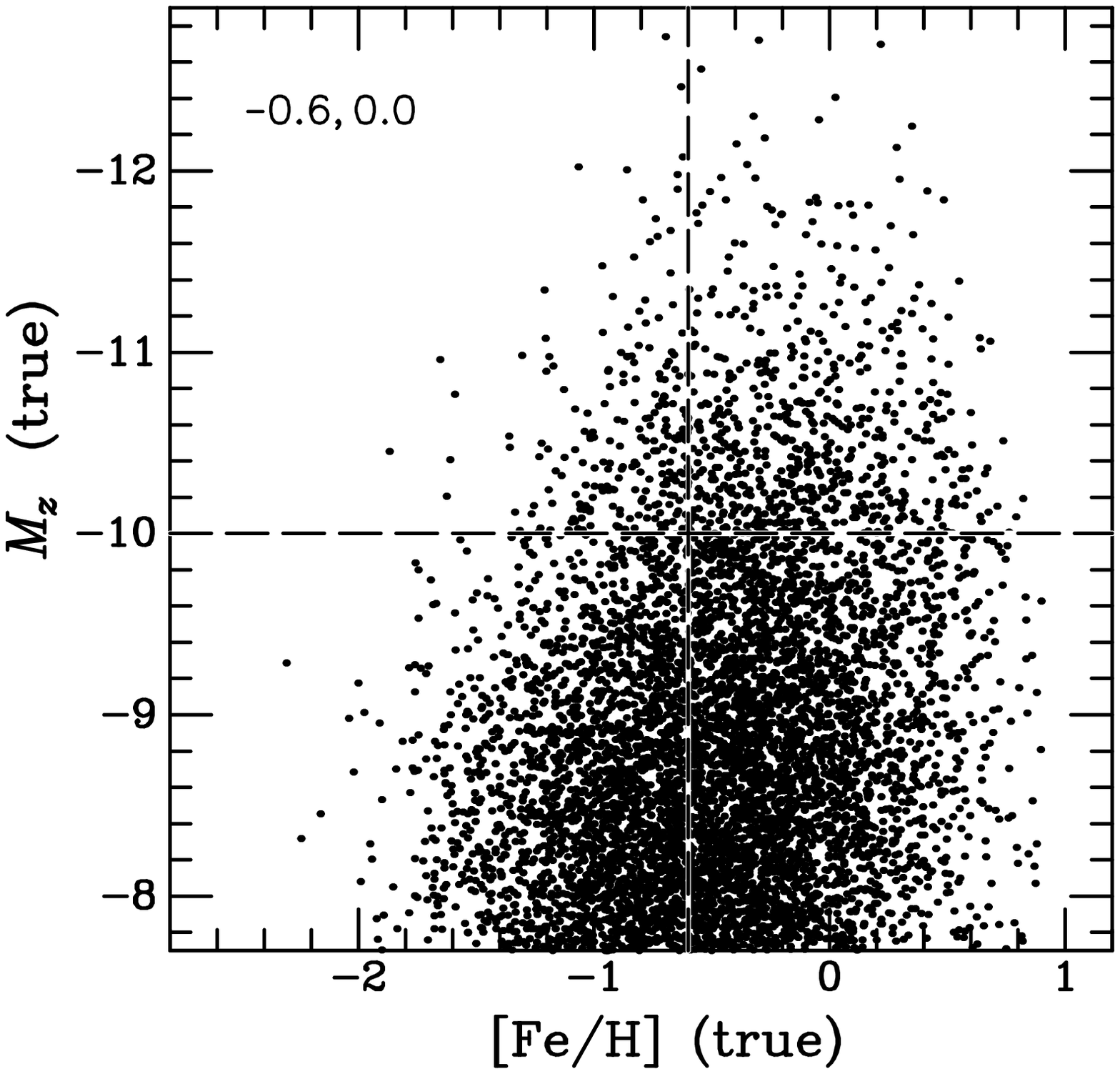}{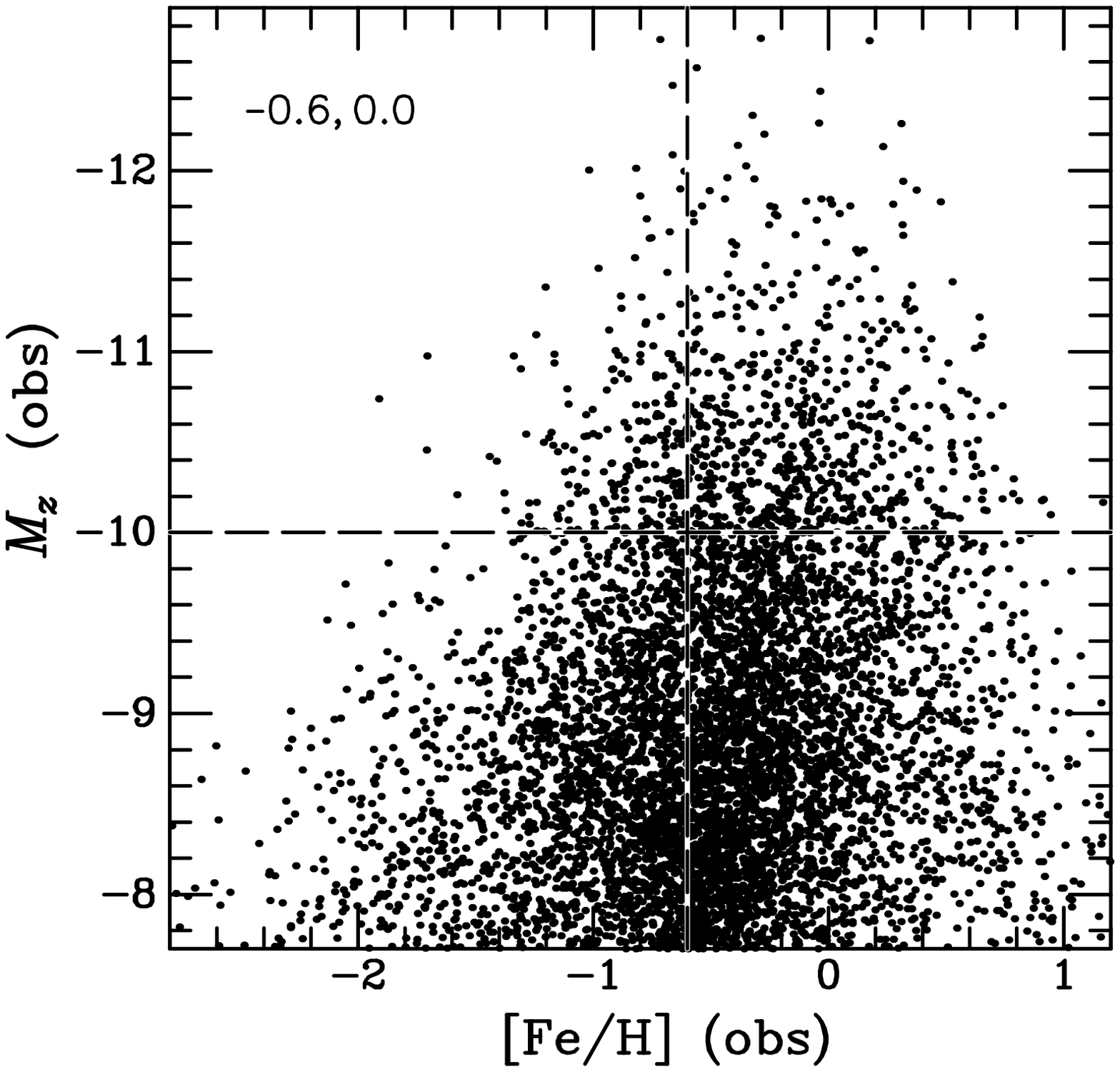}
\caption{%
Magnitude-metallicity relation
  for a simulated population of 10,000 GCs with metallicity range 
  parameters $(\feho,\fehmax) = (-0.6, 0.0)$ and $\siglf=1.4$ mag, our fiducial
  ``giant galaxy'' model.
  \textit{Left}:~true $z$-band magnitudes and
  metallicities for GCs in the simulation.
  \textit{Right}:~model ``observed'' magnitudes and the metallicities inferred
  from the \gz\ colors, including simulated observational error.  
  Consistent with empirical studies, there is no well-defined trend of 
  the observed magnitude with metallicity in the overall population;
  the right panel can be compared to Fig.~12 of Mieske \etal\ (2006), 
  to which it bears a strong resemblance. The dashed fiducial lines are the same
  as those in Mieske \etal: the horizontal line marks the approximate
  magnitude where the ``tilt'' becomes more pronounced, and the
  vertical line approximates the mean metallicity of the GCs fainter~than~this.
\label{fig:mz_feh}}
\end{figure}

For the models explored in the present work, there is no bimodality in metallicity, just
a subtle increase in the mean GC metallicity with mass, as may be expected from
self-enrichment models.  We have done this precisely to test whether or not the 
`blue tilt'' can occur within unimodal metallicity distributions.  As we have
shown, the resulting color distributions and color-magnitude relations are similar to those
found in real galaxies.  The correlation between luminosities and
metallicities in these models is dominated by scatter, as shown in Figure~\ref{fig:mz_feh}, which plots
the $z$-band magnitude-metallicity relation that would be observed for our favored giant
elliptical model.  The plotted ``observed metallicities'' in the right panel of
Figure~\ref{fig:mz_feh} are found by using the colors that include simulated
observational error and then inverting the color-metallicity relation.  The resemblance
of this figure to the corresponding one observed by Mieske \etal\ (2006, see their
Fig.~12) is striking.

Very little is actually known about the detailed metallicity distributions of 
GCs in giant ellipticals.   In support of their hypothesis, Yoon \etal\ (2006) 
showed that the histogram of Mg$b$ values for 150 GCs in M87 from Cohen \etal\ (1998) 
is reasonably consistent with a Gaussian metallicity distribution and the conversions
predicted by their SSP models; however, this represents only $1\%$ of the GCs
and may not be representative of the full population.  In contrast, Strader \etal\
(2007) found that metallicities derived from the Lick index measurements for 47 GCs in
M49 (Cohen \etal\ 2003) do appear bimodal, in the sense that a double Gaussian provided
a significantly better fit to their derived metallicities than a single Gaussian.  
%
A skewed distribution, such as found for halo stars in other galaxies
(e.g., Harris \& Harris 2002; Harris \etal\ 2007)
and derived by Yoon \etal\ (2010) from multi-band optical photometry of
extragalactic GC systems, would also favor two-Gaussian models, even
if the enrichment process was continuous. 
Perhaps more importantly, this 47-object spectroscopic sample represents only
$\sim0.6\%$ of M49's total population.
It may be that such a small sample of bright GCs observed in two spectroscopic masks
in two regions of M49 shows metallicity bimodality, but that the full population would
have a more continuous distribution from the blending of numerous substructures and
metallicity components as expected from hierarchical accretion.
%
We note that 
the metallicities as published by Cohen \etal\ for the same sample
of GCs do not strongly favor bimodality (KMM $p$-value $= 0.10$), even though they come from
the same data and correlate well with those of Strader \etal\ (2007).  This is because 
subtle changes in a few metallicities can have large effects on the
interpretation when the sample size is small.
In another study,
Puzia \etal\ (2005) presented spectroscopic metallicities, ages, and
$[\alpha/\hbox{Fe}]$ ratios for 71 GCs in seven early-type galaxies, finding
evidence for trends of decreasing GC age and $[\alpha/\hbox{Fe}]$ with increasing
metallicity.  The metallicity histogram for this sample appeared unimodal, but with
broad tails (see Fig.~7 of Puzia \etal\ 2005).  In short, published
spectroscopic data sets are inadequate for detailed constraints on the GC
metallicity distributions in the general population of early-type galaxies.

The prospects for obtaining large samples are of course more favorable with multi-band
imaging, and the combination of high-quality near-IR and optical photometry can provide 
useful constraints on metallicity (see CB07).  For instance, by combining optical and
IRAC 3.6\,$\mu$m-band photometry for 146~GCs in the peculiar, post-merger elliptical
NGC\,5128 (Cen~A), Spitler \etal\ (2008) found evidence for bimodality in metallicity.
A similar data set for 75 GCs in the Sombrero galaxy was inadequate to draw any
conclusions in this regard.  Several other important, ground-breaking studies of the optical-IR colors
of GCs in giant ellipticals have been carried out (e.g., Kissler-Patig \etal\ 2002; 
Puzia \etal\ 2002;  Hempel \etal\ 2007).
The color distributions generally appear broad, often with non-Gaussian tails, but the
samples are still relatively small and do not show the same obvious bimodality as
found in the large samples of purely optical GC colors.
A claimed exception to this was based on matched WFPC2 $I$ and NICMOS $H$ photometry
of a pruned subsample of 80 GCs in M87 (Kundu \& Zepf 2007).  The positions 
and photometry for these objects were never published, but the histogram showed a 
gap at intermediate color.  However, the red ``peak'' comprised an excess of only
4~GCs with respect to this gap, and a double Gaussian was preferred over a
single Gaussian at only the 2.4-$\sigma$ level.  Most importantly, the whole sample
constituted only 0.5\% of the full population of M87 GCs.
Representative samples of optical-IR GC colors in a statistical number of elliptical galaxies are
needed for general constraints on their metallicity distributions.
In addition, the establishment of empirical conversions from these colors to
metallicity, and detailed comparison to model predictions as a function of age, will
be essential.  Several groups are actively working on this topic, so the data required
for these efforts should be available in the near future.

\section{Conclusions}\label{sec:conclusions}

%
The combination of a realistic, nonlinear color-metallicity dependence 
with a simple mass-metallicity scaling relation motivated by self-enrichment
considerations can produce GC populations with
bimodal color distributions \textit{and} an apparent color-magnitude relation
(``blue tilt'') for the GCs within the blue peak of the color distribution.  The
underlying metallicity distributions are unimodal.  For certain metallicity ranges, the
color-magnitude relation for GCs within the red peak is weak or absent, consistent with
observations for most giant elliptical galaxies.  The models which reproduce these
features have nearly equal red and blue peaks, again consistent
with the data for giant ellipticals.  As noted by Yoon et al.\ (2006), this equality
occurs because the inflection point in the color-magnitude relation is approximately
equal to the peak metallicity of the GC system.
It is when the peak metallicity begins to move through the inflection point that the
``blue tilt'' becomes most pronounced.  As a result, the significance of the tilt
tends to increase in higher metallicity populations with redder mean colors; thus,
it would be greater in more massive galaxies and at smaller galactocentric radii,
both of which are found observationally (Mieske \etal\ 2006, 2010).

At the low-metallicity extreme of our simulations, there is actually a preference for
red tilts over blue ones because the peak of the metallicity distribution is well below
the inflection point of the color-metallicity relation.  But in practice,
the galaxies that would harbor such metal-poor systems would have low luminosities, few GCs,
and far too few red GCs to show a significant tilt.
The narrower GCLFs typical of low-luminosity galaxies also greatly reduce the
significance of any color-magnitude trends in our simulations because of the
reduced range in GC luminosity.
Further, because we have assumed a single mass-metallicity relation for all GCs, this
scenario implies that the redder, more metal-rich
GC systems of giant galaxies will have a higher mean mass
than those of dwarf ellipticals, again consistent with observations 
(\jordan\ \etal\ 2007b).

A few giant galaxies with weak or absent blue tilts such as M49 (Mieske \etal\ 2006;
Strader \etal\ 2006; but see also Lee \etal\ 2008) can be understood as simply being
consistent with stochastic variations among different model realizations.  This is
because the actual underlying metallicity distribution is broad.  If the colors linearly
reflected metallicity, so that the blue peak of the GC color distribution truly represented
a distinct subpopulation ``squeezed'' into a narrow metallicity margin that
tightly became more metal-rich with luminosity, then stochastic variations in such
a homogenous population would be small.  Thus, a large galaxy without a blue tilt 
would be truly anomalous.  However, even in our favored 
high-metallicity model that accurately reproduces the observed mean \gz\ blue and red
tilts, at least 15\% of the simulations lack significant color-magnitude slopes in the
blue peak.  Thus, it is not surprising to find a galaxy like M49, and this scenario
predicts that others will be found once large samples of giant ellipticals with
similarly well-measured color-magnitude diagrams are available.

%
Further  observations are needed to test
this scenario more quantitatively.  In particular, we still do not know in detail the 
underlying form of the metallicity distributions of GCs in giant ellipticals, 
and we have assumed Gaussians only for the sake of simplicity, 
whereas the true distributions may be skewed 
in a manner similar to that of old halo stars (Yoon \etal\ 2010).  Accurate
characterization of these distributions will require still larger spectroscopic
samples and improved models for converting measured indices to actual metallicities.
As we have noted, large
samples of optical-IR colors in multiple elliptical galaxies, and empirical
conversions from these colors to metallicity, are also needed.  
The required data should be available soon from ongoing observational efforts.

We have found that the precise shapes of GC color distributions can depend
on the fine details of the color-metallicity conversion.  
%
In addition to more extensive data samples,
improving the accuracy of photometrically estimated GC metallicities will require
a broader and more densely calculated set of models,
all with better calibrations against resolved stellar populations.  The variation of
alpha-element enhancement as a function of metallicity needs to be calibrated and
taken into account for accurate data-model comparisons.   Further
refinement in our understanding of changes in the color-metallicity relation with
age, particularly with regards to the behavior of the horizontal branch, is essential.
It is also important to obtain useful
empirical constraints on trends in GC age with metallicity, although there
will likely be
large variations in any such trends due to the merging histories of individual galaxies.
A still more challenging task is to measure the change in color distributions with
age by obtaining high quality photometry of GC populations at significant cosmological
lookback times.

Finally, we note that this investigation has explored the behavior of the blue
tilt phenomenon under variations of only two parameters, here labeled \feho\ and
\fehmax.  Although we consider the results presented here promising, more
sophisticated models are of course desirable, especially once more precise
color-metallicity transformations become available.  Given the remaining
uncertainties and the importance of GC metallicity distributions to understanding
detailed galaxy formation histories, we hope that this will continue to be an
active area of theoretical and observational research
into the coming era of \textit{JWST} and the planned ELTs.

\acknowledgements  
We thank the anonymous referee for helpful comments.
It is a pleasure to thank S.-J.\ Yoon and P.~C{\^o}t{\'e} for
enlight\-ening discus\-sions.
M.C.\ acknowledges support from ASI-INAF (program I/016/07/0), PRIN-INAF
2006 (P.I. G.\ Clementini), and PRIN-INAF 2008 (P.I. M.\ Marconi).
E.W.P.\ gratefully acknowledges the support of the Peking
University Hundred Talent Fund (985).
J.P.B.~thanks \hbox{N.-J.~Kim} for providing \hbox{inspiration} for this work.

\end{document}